\documentclass[prl, twocolumn, reprint]{revtex4-1}

\usepackage{amsmath, graphicx, amsthm, natbib}
\usepackage[utf8]{inputenc}
\usepackage[english]{babel}

\newcommand{\ket}[1] {|#1\rangle}
\newcommand{\bra}[1] {\langle #1|}
\newcommand{\overlap}[2]{\langle #1|#2\rangle}
\newcommand{\tr}{{\rm Tr}}
\newcommand{\hH}{\hat{\mathcal{H}}}
\newcommand{\hs}{\hat\sigma}
\newcommand{\hU}{\hat{U}}
\newcommand{\hW}{\hat{W}}
\newcommand{\hV}{\hat{V}}
\newcommand{\+}{^{\dagger}}

\begin{document}
\title{Proposal to measure out-of-time-ordered correlations using Bell states}
\author{Bhuvanesh Sundar}
\email{bhuvanesh.sundar@uibk.ac.at}
\affiliation{Institute for Quantum Optics and Quantum Information of the Austrian Academy of Sciences, Innsbruck A-6020, Austria}

\begin{abstract}
We present a protocol to experimentally measure the infinite-temperature out-of-time-ordered correlation (OTOC) -- which is a probe of quantum information scrambling in a system -- for systems with a Hamiltonian which has either a chiral symmetry or a particle-hole symmetry. We show that the OTOC can be obtained by preparing two entangled systems, evolving them with the Hamiltonian, and measuring appropriate local observables. At the cost of requiring two copies of the system and putting restrictions on the Hamiltonian's symmetries, we show that our method provides some advantages over existing methods -- it can be implemented without reversing the sign of the Hamiltonian, it requires fewer measurements than schemes based on implementing the SWAP operator, and it is robust to imperfections like some earlier methods. Our ideas can be implemented in currently available quantum platforms.
\end{abstract}

\maketitle

{\em Introduction}.--
Quantum information scrambling studies the spreading of initially local information through a quantum system. Information scrambling is deeply connected to fundamental concepts in physics such as quantum chaos~\cite{zhu2016measurement, yao2016interferometric, belyanski2020minimal, zhang2019information, nahum2018operator, hosur2016chaos, roberts2017chaos, syzranov2019interaction, belyanski2020minimal}, localization~\cite{fan2017out, chen2017out, syzranov2019interaction, he2017characterizing, slagle2017out, huang2017out, swingle2017slow, chen2016universal}, phase transitions~\cite{daug2019detection, wei2019dynamical, sun2018out, wang2019probing, heyl2018detecting, shen2017out, nie2019experimental, swan2020diagnosing}, and thermodynamics~\cite{tsuji2018out, campisi2017thermodynamics, halpern2017jarzynski, halpern2018quasiprobability, swan2020diagnosing}, and finds applications in studies of black holes~\cite{maldacena2016bound, hayden2007black, shenker2014black} and condensed matter models with holographic duals~\cite{sachdev1993gapless, sachdev2010holographic, banerjee2017solvable, kitaev2017simple, danshita2017creating, chew2017approximating, gu2017local, chen2018quantum}.

The scrambling of quantum information can be probed by measuring the squared magnitude of the commutator between two local observables at different times, $C(t) = \langle [\hW,\hV(t)]\+ [\hW,\hV(t)]\rangle$. This quantity probes the spreading of the Heisenberg operator $\hV(t)$ by giving the noncommutativity of $\hW$ with $\hV(t)$. In chaotic systems, $C(t)$ exhibits a period of exponential growth, $C(t)\sim e^{\lambda_L t}$, where $\lambda_L$ is bounded by an upper limit of $2\pi k_BT/\hbar$ ~\cite{maldacena2016bound, hayden2007black, shenker2014black}. When $C(t)$ is expanded, it contains time-ordered correlations, $\langle \hW\+\hV\+(t)\hV(t)\hW\rangle$ and $\langle \hV\+(t)\hW\+\hW\hV(t)\rangle$, and out-of-time ordered correlations (OTOCs), $\langle \hW\+\hV\+(t)\hW\hV(t)\rangle$ and $\langle \hV\+(t)\hW\+\hV(t)\hW\rangle$.

Experimentally measuring OTOCs has proven to be difficult, since the order of operators in OTOCs suggests that sign-reversal of the Hamiltonian is required. Researchers have proposed to measure OTOCs by explicitly reversing the sign of the Hamiltonian~\cite{swingle2016measuring}, or by controlling the sign with an ancillary bit which acts as a switch~\cite{zhu2016measurement}. OTOCs have been measured by explicitly reversing the Hamiltonian's sign in NMR quantum simulators~\cite{li2017measuring, wei2018exploring, nie2019detecting} and a system with trapped ions~\cite{garttner2017measuring}. Other proposals to measure OTOCs without reversing time evolution involve implementing the SWAP operator between two systems either as an ensemble of random initial states~\cite{vermersch2019probing} or using a beam splitter operation~\cite{yao2016interferometric}, or making weak measurements~\cite{halpern2017jarzynski}. A landmark experiment~\cite{joshi2020quantum} recently measured OTOCs by implementing the proposal in Ref.~\cite{vermersch2019probing}. OTOCs have also been measured in an experimental implementation~\cite{landsman2019verified} of the Hayden-Preskill protocol~\cite{hayden2007black, yoshida2017efficient, yoshida2019disentangling}.

In this Letter, we propose a method to measure OTOCs at infinite temperature, $\langle \hW\+\hV\+(t)\hW\hV(t)\rangle_\infty$ and $\langle \hV\+(t)\hW\+\hV(t)\hW\rangle_\infty$, for Hamiltonians which have a chiral symmetry or a particle-hole symmetry. Our method works by measuring quantum correlations between two systems that are initially entangled and then evolved with the Hamiltonian. The condition on the Hamiltonian's symmetry arises from a special property of our initial state, which for these symmetries, effectively evolves one of the systems backward in time without requiring to reverse the Hamiltonian's sign in experiment. At the cost of requiring two copies of the system and restricting to Hamiltonians with certain symmetries, we show that our scheme provides some advantages over earlier methods that measure OTOCs. First, as mentioned above, our scheme does not require reversing the Hamiltonian's sign, which is a significant advantage over methods which reverse the  sign~\cite{swingle2016measuring, zhu2016measurement, li2017measuring, wei2018exploring, nie2019detecting, garttner2017measuring}. Second, it requires fewer measurements than methods which measure the SWAP operator~\cite{yao2016interferometric, vermersch2019probing, joshi2020quantum}. Additionally, like earlier works~\cite{vermersch2019probing, joshi2020quantum}, our method is also robust to imperfections in experiment.
 
 We demonstrate our method by applying it to measure OTOCs for two Pauli operators in a system with a non-integrable spin Hamiltonian which naturally arises in Rydberg systems~\cite{de2019observation}. For this case, we initially entangle the qubits in two systems as Bell pairs, apply a Pauli operator on one system, evolve both systems with the Hamiltonian, and then measure the correlation between a Pauli operator in the two systems. This scheme is readily implementable in current experiments. For general OTOCs beyond Pauli operators, the two systems in our proposal have to be initially entangled in such a way that they form a purified state of a certain operator. We present a variationally-inspired algorithm to prepare this initial state. We focus on qubits, but our ideas can be applied to systems with other local Hilbert spaces too.

\begin{figure}[t]
\includegraphics[width=0.85\columnwidth]{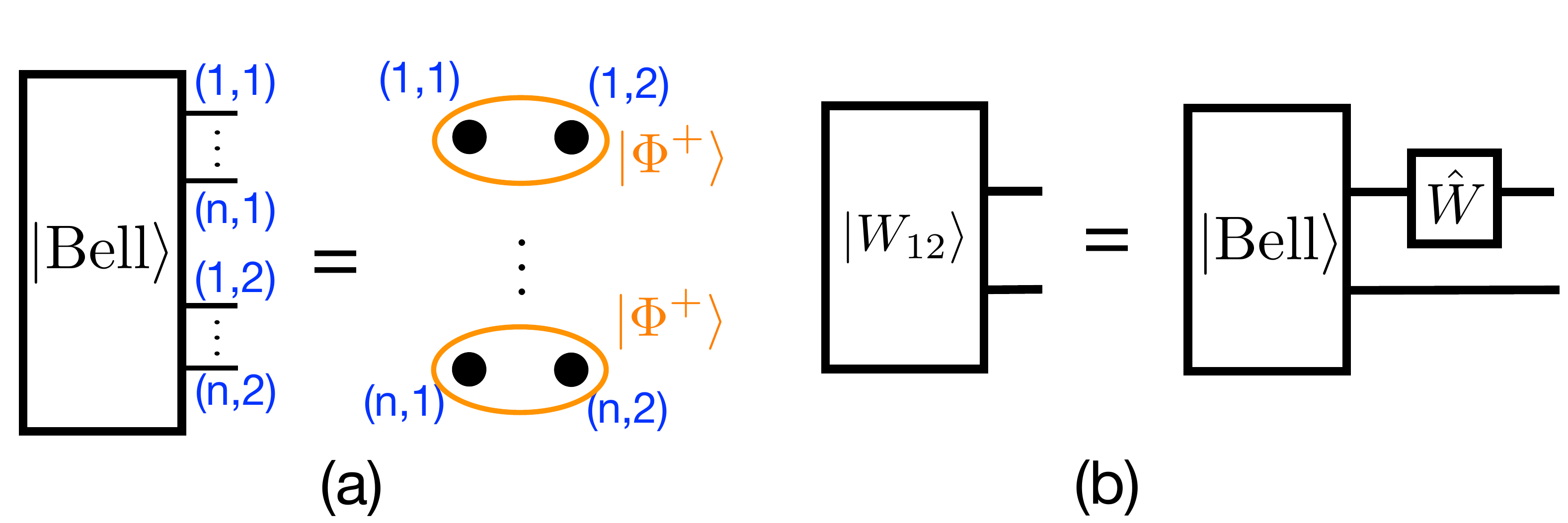}
\caption{(Color online) (a) Illustration of $\ket{\rm Bell}$ as a product of Bell pairs, when the many-body basis states are products of the single-qubit basis states $\{\ket{0},\ket{1}\}$. (b) Quantum circuit to prepare $\ket{W_{12}}$ for unitary $\hW$. }
\label{fig1: definition}
\end{figure}

\begin{figure}[t]
\includegraphics[width=1.0\columnwidth]{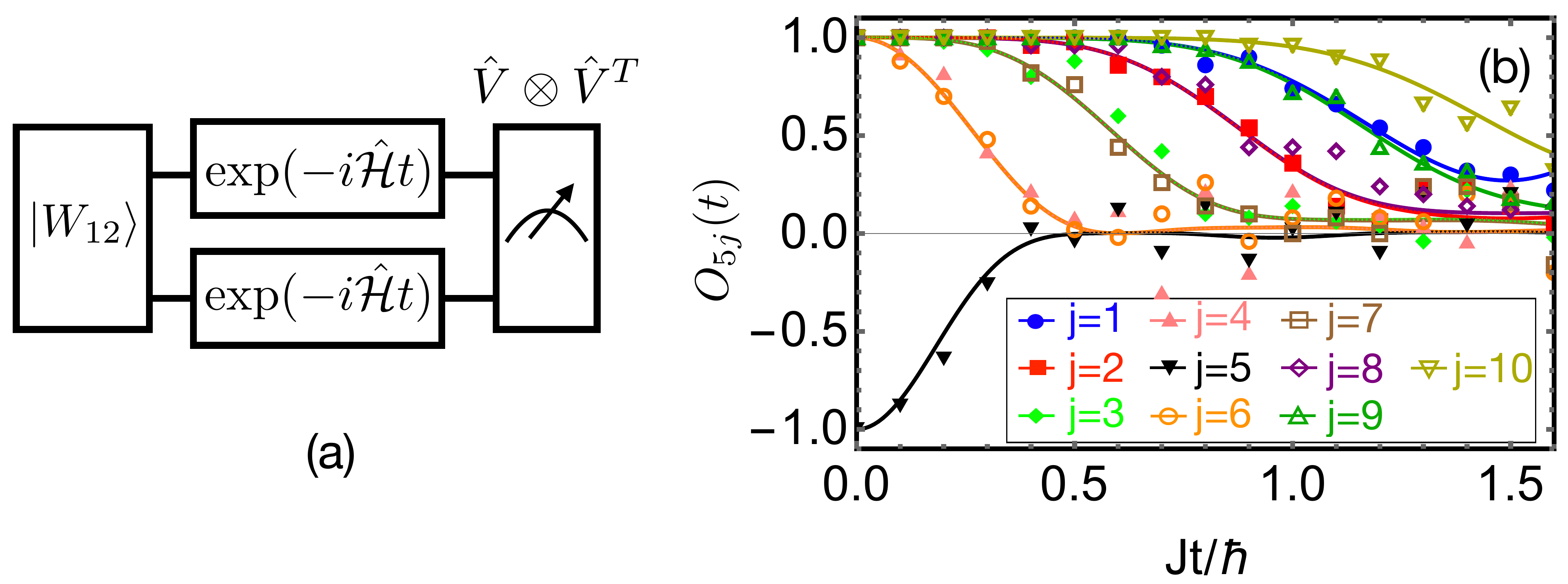}
\caption{(Color online) (a) Quantum circuit to measure the OTOC, for Hermitian $\hV$. (b) Infinite-temperature OTOCs $O_{5j}(t) = \langle \hs^z_5\hs^x_j(t)\hs^z_5\hs^x_j(t)\rangle_\infty$ for the Hamiltonian in Eq.~\eqref{eqn: H ssh} for $n=10$ qubits. Solid lines show the exact values, while the symbols show the results of a simulated experiment with $100$ measurements for each time. Deviations of the symbols from the solid lines are due to shot noise.}
\label{fig2: otoc}
\end{figure}

{\em Measuring the OTOC for unitary $\hW$}.--
We prepare the two systems initially in 
$
\ket{W_{12}} = \left( \hW \otimes \hat{1} \right) \ket{{\rm Bell}},
$ 
where
\begin{equation}\label{eqn: bell}
\ket{{\rm Bell}} = \frac{1}{2^n} \sum_{\ket{x}} \ket{x}_1\ket{x}_2.
\end{equation}
The sum in Eq.~\eqref{eqn: bell} runs over basis states $\{\ket{x}\}$ that will be chosen later, and the subscripts label the two systems. $\ket{{\rm Bell}}$ can be prepared relatively easily for simple choices of $\{\ket{x}\}$. For example, if the many-body basis states are products of single-qubit states $\{\ket{0},\ket{1}\}$, then $\ket{{\rm Bell}} = \otimes_{j=1}^n \ket{\Phi^+_j}$ is a product of Bell pairs as illustrated in Fig.~\ref{fig1: definition}(a), with $\ket{\Phi^\pm_j} = (\ket{0}_{(j,1)}\ket{0}_{(j,2)} \pm \ket{1}_{(j,1)}\ket{1}_{(j,2)})/\sqrt{2}$. This state can be prepared relatively easily on most experimental platforms that perform quantum simulation. Then, $\ket{W_{12}}$ can be prepared by applying $\left( \hW\otimes\hat{1} \right)$ to $\ket{{\rm Bell}}$, as shown in Fig.~\ref{fig1: definition}(b).

Our proposal to measure the OTOC stems from the relation
\begin{align}\label{eqn: corollary}
\langle W_{12}(t)| & \left(\hV\+\otimes\hV^T\right) |W_{12}(t)\rangle = \frac{ 1 } { 2^n } \times \nonumber\\ &\tr \bigg( \hW\+ e^{i\hH t}\hV\+ e^{-i\hH t}\hW e^{-i\hH^Tt}\hV e^{i\hH^Tt}\bigg),
\end{align}
where $\ket{W_{12}(t)} = \left(e^{-i\hH t}\otimes e^{-i\hH t}\right) \ket{W_{12}}$, and we used $\langle x|\hat{V}^T|x'\rangle = \langle x'|\hat{V}|x\rangle$ to derive Eq.~\eqref{eqn: corollary}. We set $\hbar=1$.

Equation~\eqref{eqn: corollary} gives the infinite-temperature OTOC $\langle \hW\+\hV\+(t)\hW\hV(t)\rangle_\infty$, if the Hamiltonian satisfies $\hH^T=-\hH$. The circuit to measure the OTOC for Hermitian $\hV$ is shown in Fig.~\ref{fig2: otoc}(a). We highlight that this circuit evolves both systems with $+\hH$. For Hermitian $\hV$, $\langle \hW\+\hV\+(t)\hW\hV(t)\rangle_\infty = \langle \hV\+(t)\hW\+\hV(t)\hW\rangle_\infty$, and both can be obtained by measuring $\hV\otimes\hV^T$ in $\ket{W_{12}(t)}$. 

Two key points explain why our protocol can measure the OTOC without reversing the sign of $\hH$. First, $\ket{{\rm Bell}}$ is an isotropic state, which satisfies
\begin{equation}\label{eqn: purified states rotation corollary}
\left(\hU\otimes\hU^*\right)\ket{{\rm Bell}} = \ket{{\rm Bell}}
\end{equation}
for any unitary $\hU$. Second, $\left(e^{-i\hH t}\right)^* = e^{-i\hH t}$ if $\hH^T=-\hH$. Then, setting $\hU=e^{-i\hH t}$ in Eq.~\eqref{eqn: purified states rotation corollary}, and multiplying Eq.~\eqref{eqn: purified states rotation corollary} by $(e^{i\hH t}\otimes\hat{1})$, we find $\left( \hat{1}\otimes e^{-i\hH t}\right)\ket{{\rm Bell}} = \left( e^{i\hH t}\otimes \hat{1}\right)\ket{{\rm Bell}}$. That is, we \textit{effectively} evolve system 1 with $-\hH$, by evolving system 2 with $+\hH$.

The requirement $\hH^T=-\hH$ is satisfied in \textit{some} basis for all $\hH$ with either a chiral symmetry or a particle-hole symmetry. Then to measure the OTOC, one uses this basis to define $\ket{{\rm Bell}}$ [Eq.~\eqref{eqn: bell}], and implements the circuit in Fig.~\ref{fig2: otoc}(a). While this restricts the applicability of our method, it still lets us measure the OTOC for several $\hH$ describing a large class of physical systems. The biggest challenge in our protocol is finding a basis where $\hH^T=-\hH$ and $\ket{{\rm Bell}}$ can be prepared in experiment.

We demonstrate our method by applying it to calculate the OTOC for the Hamiltonian
\begin{equation}\label{eqn: H ssh}
\hH_{AB} = \sum_{ij} \frac{J}{r_{ij}^3} \left(\hs^x_{A,i}\hs^x_{B,j} + \hs^y_{A,i}\hs^y_{B,j} \right),
\end{equation}
where $(A,i)$ and $(B,j)$ denote qubits $i$ and $j$ on $A$ and $B$ sublattices of a 1D chain. This Hamiltonian is non-integrable, has a chiral symmetry, and a close variant of it has been realized in recent experiments on Rydberg atoms~\cite{de2019observation}. The chiral symmetry in these experiments arises when the atoms' dipole moment is aligned at an angle of $\cos^{-1}(1/\sqrt{3})$ with respect to the two legs of a $2\times (n/2)$ ladder of atoms.

Figure~\ref{fig2: otoc}(b) plots the OTOCs, $O_{5j}(t) = \langle \hs^z_5\hs^x_j(t)\hs^z_5\hs^x_j(t)\rangle_\infty$, for $\hH_{AB}$ on a chain of $n=10$ qubits. $\hH_{AB}$ has an anti-symmetric matrix when the basis states are chosen as $\{\ket{0}_j, \ket{1}_j\}$ for even-numbered qubits $j=2k$, and $\{\ket{0}_j, i\ket{1}_j\}$ for odd-numbered qubits $j=2k+1$, $k\in Z$. For this basis, $\ket{W_{12}} = \left( \otimes_{j\in{\rm odd},j\neq5} \ket{\Phi^-_j} \right)\left( \otimes_{j\in{\rm even} || j=5} \ket{\Phi^+_j} \right)$ is also a product of Bell pairs, and can be prepared in experiments. Figure~\ref{fig2: otoc}(b) shows that all the off-site ($j\neq5$) OTOCs are initially $1$, since the operators initially commute, and the on-site ($j=5$) OTOC is $-1$ since the operators initially anti-commute. The OTOCs begin decaying with time, with the onset of decay happening at a later time when the initial operators are spaced farther apart.

{\em Comparison to earlier methods}.-- Our work shares some aspects with earlier proposals~\cite{vermersch2019probing} and experiments~\cite{joshi2020quantum, landsman2019verified} that measured OTOCs, which we now describe.

The essence of Refs.~\cite{vermersch2019probing, joshi2020quantum} is the relation
\begin{equation}\label{eqn: random unitaries}
\overline{ a_k \left(\hat{u}\otimes \hat{u}\right) \ket{ 0^{\otimes n}\otimes k } \bra{ 0^{\otimes n}\otimes k } \left(\hat{u}\otimes \hat{u}\right)^\dagger } =  \frac{ \alpha \hat{1} + \beta {\rm\ SWAP} }{ 2^n }
\end{equation}
for appropriately chosen weights $a_k$. Equation~\eqref{eqn: random unitaries} gives the density matrix for their ensemble of initial states, up to a normalization constant. The average $\overline{\cdots}$ is over a set of random unitaries $u$ which are either local or global, and a set of initial bit strings $k$. In the protocol with global random unitaries, $a_k=\delta_{k,0}$, and $\alpha = \beta = 1/(2^n+1)$. In the protocol with local random unitaries, Ref.~\cite{vermersch2019probing} chose $a_k = (-2)^{-D[k]}$ with $D[k]$ the Hamming weight of $k$, and proposed a converging series to estimate the OTOC. If $k$ is averaged over all $2^n$ bit strings, then $\alpha=0$ and $\beta = 2^{-n}$. If $k$ is instead restricted to only $2^L$ bit strings, varying $L$ qubits close to the location of $\hV(0)$ and leaving all other qubits as $\ket{0}$, then Eq.~\eqref{eqn: random unitaries} is modified to a product of local density matrices $(\alpha_i \hat{1} + \beta_i {\rm\ SWAP}_i)/2$, with $\alpha_i=0,\beta_i=1/2$ for the $L$ qubits, and $\alpha_i=\beta_i=1/3$ for the others. The total number of measurements required by these protocols roughly scales as $2^n$ for the first two cases, and $2^L$ for the last case.

Refs.~\cite{vermersch2019probing, joshi2020quantum} differ from our method only in the initial state, and have an identical circuit otherwise. Their initial state [Eq.~\eqref{eqn: random unitaries}] is an eigenstate of $\hU\otimes\hU$ for any unitary $\hU$. Therefore, system 1 \textit{effectively} evolves with $-\hH$ when system 2 is evolved with $+\hH$, without any restrictions on the symmetries of $\hH$. Moreover, the two systems can be simulated in separate experiments with only $n$ qubits each, and the measurements can be classically correlated. The method in Refs.~\cite{vermersch2019probing, joshi2020quantum} is advantageous over our method in these two respects. However, as we will show, our method has the advantage that it requires fewer measurements. This is because our initial density matrix is $\ket{{\rm Bell}}\bra{\rm Bell}$, with the prefactor $\beta=1$.

The identity in Eq.~\eqref{eqn: purified states rotation corollary} plays a key role in the Hayden-Preskill protocol~\cite{hayden2007black, yoshida2017efficient, yoshida2019disentangling}. Ref.~\cite{landsman2019verified} demonstrated this protocol in experiment by applying $U\otimes U^*$ to a system consisting of Bell pairs and an unknown state $\ket{\psi}$. The operations $U\otimes U^*$ scramble $\ket{\psi}$ across the system, which is then recovered elsewhere by measuring a small number of qubits in the Bell basis. Our method uses Eq.~\eqref{eqn: purified states rotation corollary} to measure OTOCs, and uses the fact that $U^*=U$ for Hamiltonians with an anti-symmetric matrix. We also show how to use Eq.~\eqref{eqn: purified states rotation corollary} to detect errors in the system.

\begin{figure}[t]\centering
\includegraphics[width=1.0\columnwidth]{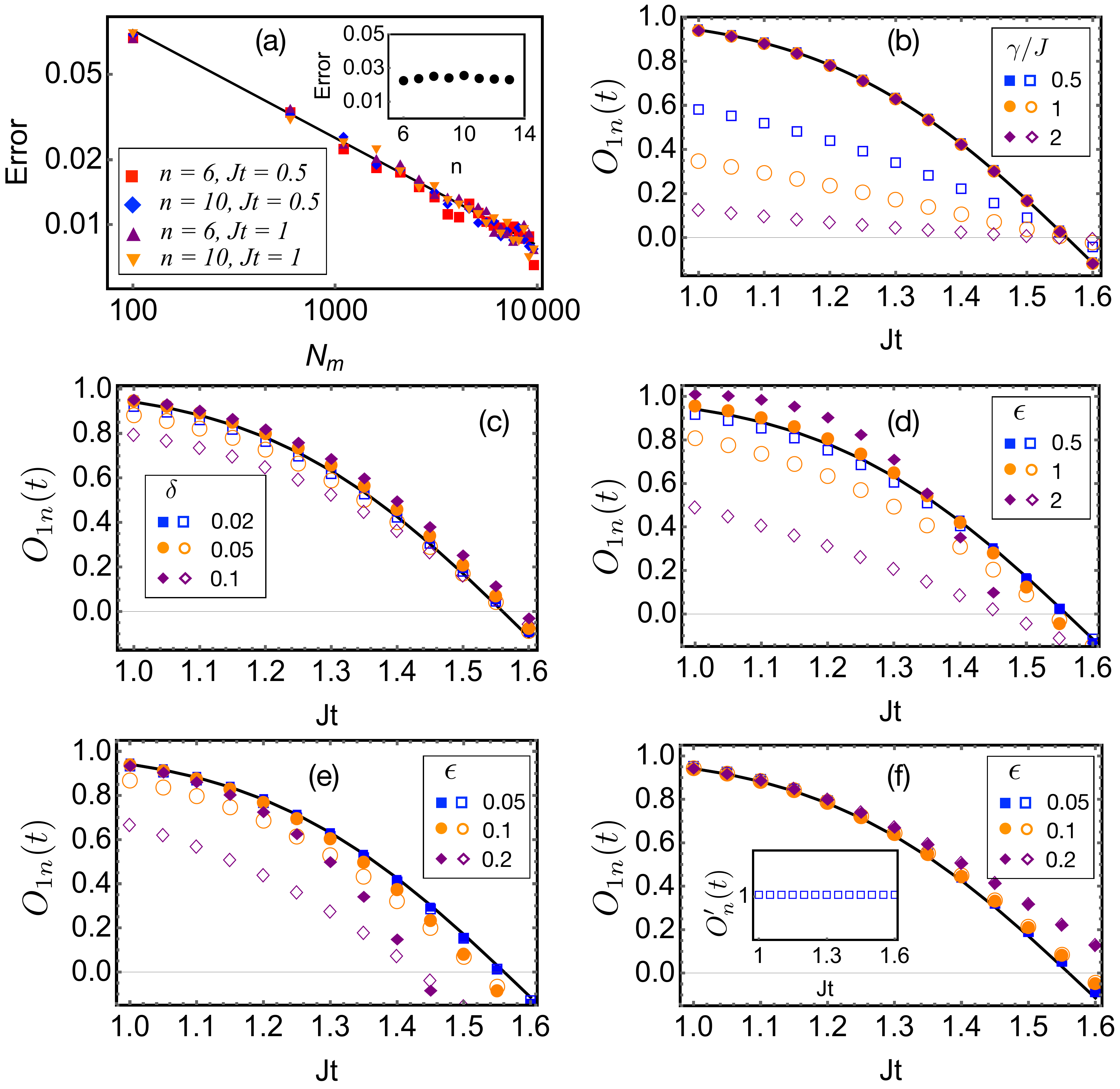}
\caption{(Color online) Statistical errors, imperfections, and decoherence. (a) The statistical error in the estimated OTOC $O_{n/2,n/2+1}^{\rm est}(t)$ due to shot noise, as a function of the number of measurements $N_m$. The solid line shows $1/\sqrt{N_m}$, and the symbols are the error for two different system sizes and times. The inset plots the error for $O_{n/2,n/2+1}^{\rm est}(t)$ at $Jt=0.5$ for $N_m=1000$ measurements, and shows that the error stays constant with system size. (b-f) Open symbols: Estimated OTOC $O_{1n}^{\rm est}(t)$ in the presence of errors, for a chain of $n=6$ qubits. Filled symbols: Rescaled OTOC $\overline{O}_{1n}(t) = O_{1n}^{\rm est}(t)/O'_n(t)$. (b) includes depolarizing noise. (c) considers imperfect preparation of $\ket{\rm Bell}$. (d) includes chiral-symmetry-breaking terms in the Hamiltonian. (e) considers the two systems to evolve with different Hamiltonians. (f) includes coupling between the two systems. Inset in (f) shows that $O'_n(t) = 1$ for this coupling.}
\label{fig3: errors}
\end{figure}

{\em Statistical errors, imperfections, and decoherence}.--
Figure~\ref{fig3: errors}(a) shows the statistical error in the measured value of $O_{ij}(t) = \langle \hs^z_i\hs^x_j(t)\hs^z_i\hs^x_j(t) \rangle_\infty$ for the two central qubits in a chain, as a function of the number of measurements $N_m$, for two different system sizes and evolution times. The shot noise decreases as $\sqrt{N_m}$, and does not increase with system size or evolution time.

In addition to statistical error, we expect errors to occur in an experimental implementation of our proposal due to imperfect initial state preparation, readout errors, symmetry-breaking terms in the Hamiltonian, coupling between the two systems, unequal Hamiltonians in the two systems, and other decohering processes such as depolarizing noise and spontaneous emission. We propose to detect these errors by measuring $O'_j(t) = \langle \hs^x_j(t)\hs^x_j(t)\rangle_\infty$, obtained by initializing the two systems in $\ket{{\rm Bell}}$ and measuring $\hs^x_j\otimes \left(\hs^x_j\right)^T$ after time evolution. $\ket{{\rm Bell}}$ is an eigenstate of $\hs^x_j\otimes \left(\hs^x_j\right)^T$, and an eigenstate of $\left( e^{-i\hH t}\otimes e^{-i\hH t} \right)$ if $\hH^T=-\hH$. Therefore, in the ideal case of no errors, measuring $\hs^x_j\otimes \left(\hs^x_j\right)^T$ should always yield $1$. Any deviation from $1$ indicates that an error has occurred. Dividing $O_{ij}(t)$ by $O'_j(t)$ removes some of these errors, as we explain below.

Depolarizing noise in the experiment produces smaller estimates $O_{ij}^{\rm est}(t)$ and $O'_j(t)$ than the ideal results $O_{ij}(t)$ and $1$. For depolarization rate $\gamma$, $O_{ij}^{\rm est}(t) = e^{-\gamma t} O_{ij}(t)$ and $O'_j(t) = e^{-\gamma t}$. This error is completely eliminated by calculating the ratio $\overline{O}_{ij}(t) = O_{ij}^{\rm est}(t)/O'_j(t)$, which recovers $O_{ij}(t)$ exactly. Readout errors are similarly cancelled in $\overline{O}_{ij}(t)$ (see Supplementary Material).

Figure~\ref{fig3: errors}(b) shows our method's robustness to depolarizing noise. The open symbols plot $O_{1n}^{\rm est}(t)$ and the filled symbols plot $\overline{O}_{1n}(t)$. The filled symbols overlap with the exact result $O_{1n}(t)$ (solid line). Then, experiments can accurately extract $O_{1n}(t)$ as long as $O_{1n}^{\rm est}(t)$ is above a shot noise threshold set by the number of measurements.

Figures~\ref{fig3: errors}(c-f) plot $O_{1n}^{\rm est}(t)$ and $\overline{O}_{1n}(t)$ in the presence of other sources of error. Figure~\ref{fig3: errors}(c) considers the initial density matrix to be $\hat{\rho}_{\rm init} = \otimes_j ((1-\delta) \hat{\rho}_j + \hat{1}\times\delta/4)$, where $\hat{\rho}_j$ is the ideal Bell state for the $j^{\rm th}$ qubit. Figure~\ref{fig3: errors}(d) considers the Hamiltonian in each system to be $\hH = \hH_{AB} + \epsilon(\hH_{AA}+\hH_{BB})$, where the chiral-symmetry-breaking terms $\hH_{AA}$ and $\hH_{BB}$ are spin interactions within a sublattice that decay as $1/r^3$. Figure~\ref{fig3: errors}(e) considers the two systems to evolve with different Hamiltonians, $\hH_{1(2)} = (1+\epsilon_{1(2)}) \hH_{AB}$ with $\epsilon_{1(2)} = +\epsilon (-\epsilon)$. Figure~\ref{fig3: errors}(f) considers additional coupling between the two systems, $\hH_{\rm coupling} = \epsilon \sum_i J(\hs^x_{i1}\hs^x_{i2} + \hs^y_{i1}\hs^y_{i2})$. The rescaled value, $\overline{O}_{1n}(t)$, overestimates $O_{1n}(t)$ in the case of imperfect initial states [Fig.~\ref{fig3: errors}(c)] and when the systems are coupled [Fig.~\ref{fig3: errors}(f)]. When symmetry-breaking terms are present [Fig.~\ref{fig3: errors}(d)], $\overline{O}_{1n}(t)$ overestimates $O_{1n}(t)$ at the onset of the decay and underestimates $O_{1n}(t)$ at later times. $\overline{O}_{1n}(t)$ underestimates $O_{1n}(t)$ when the systems evolve with different Hamiltonians [Fig.~\ref{fig3: errors}(e)]. Moreover, $O'_n(t)=1$ in this case [inset in Fig.~\ref{fig3: errors}(f)], so this coupling will not be detected by $O'_n(t)$. We consider spontaneous emission in the Supplementary Material.

{\em Preparing $\ket{W_{12}}$ for non-unitary $\hW$}.--
The OTOC for non-unitary $\hW$ can also be measured from Eq.~\eqref{eqn: corollary} and Fig.~\ref{fig2: otoc}(a), with the initial state $\ket{W_{12}}$ still defined the same as before and normalized. The normalized state is 
\begin{align}\label{eqn: W12}
\ket{W_{12}} =& \sqrt{\frac{2^n}{\tr(\hW\hW\+)}} \left(\hW\otimes \hat{1}\right) \ket{{\rm Bell}}\nonumber\\ =& \frac{1}{\sqrt{\tr(\hW\hW\+)}} \sum_{\ket{w}} w \ket{w}_1\ket{w^*}_2,
\end{align}
where the sum runs over the eigenstates $\ket{w}$ of $\hW$, with $w$ the corresponding eigenvalue for $\ket{w}$. The complex conjugate $\ket{w^*}$ is defined as $\ket{w^*} = \sum_{\ket{x}} \ket{x}\overlap{w}{x}$. Equation~\eqref{eqn: W12} can be derived using $\hW = \sum_{\ket{w}} w\ket{w}\bra{w}$. The state $\ket{W_{12}}$ is a purified state of $\hW\hW\+$.

Ref.~\cite{sels2019quantum} presented a probabilistic protocol to experimentally prepare $\ket{W_{12}}$ for non-unitary $\hW$. Their protocol required post-selection on a control qubit, and its success decreased as the fidelity increased.

Here, we present a deterministic protocol to coherently prepare $\ket{W_{12}}$ for $\hW$ that is non-unitary, Hermitian and easily diagonalizable. This requirement is not severely limiting, since most observables of interest are easily diagonalizable. Our protocol is inspired by a striking similarity between the symmetry of $\ket{W_{12}}$ and that of the wave function that appears in Grover's algorithm~\cite{grover1996fast}.

We denote the qubits where $\hW$ has support to be $[1, k]$, and assume for simplicity that the many-body basis states are products of single-qubit states $\{\ket{0},\ket{1}\}$. We define $U_x = 1-2\ket{\psi_0}\bra{\psi_0}$ as the reflection operator about $\ket{\psi_0} = \otimes_{i=1}^k \ket{+_i}$, and define
\begin{equation}\label{eqn: W1}
\ket{W_1} = \sum_w w \ket{w}_1.
\end{equation}

We show how to prepare $\ket{W_1}$ for diagonal $\hW$. The eigenstates $\ket{w}$ in this case are bit strings. Therefore, after preparing $\ket{W_1}$, $\ket{W_{12}}$ can be obtained by applying a controlled-NOT between every qubit in system 1 and the corresponding qubit in system 2. For $\hW$ that is not diagonal but is related to a diagonal observable $\hW^{\rm diag}$ via a unitary transformation, $\hW = \hU_W \hW^{\rm diag} \hU_W\+$, we prepare $\ket{W^{\rm diag}_{12}}$ using the method below, and then prepare $\ket{W_{12}}$ using $\ket{W_{12}} = \left(\hU_W \otimes \hU_W^*\right) \ket{W^{\rm diag}_{12}}$.

Our proposal to prepare $\ket{W_1}$ for diagonal $\hW$ relies on~\cite{sundar2019quantum, grover2020baertschi} (also see Supplementary Material)
\begin{equation} \label{eqn: psivar}
\ket{\psi_{\rm var}(\alpha_1\cdots\alpha_p)} = \prod_{j=1}^p U_x e^{i\alpha_j\hW} \ket{+^{\otimes n}} = \sum_w f(w)\ket{w}
\end{equation}
for some function $f$, for any values of $\alpha_j$. The important result in Eq.~\eqref{eqn: psivar} is that all degenerate eigenstates $\ket{w}$ with the {\em same} eigenvalue $w$ have the same coefficient. Grover's algorithm is a special case of Eq.~\eqref{eqn: psivar}, with $\alpha_j=\pi$ and $\hW$ an oracle with only two distinct eigenvalues, $w=0$ and $w=1$. Then, $f(0)=\cos((2p+1)\theta)/\sqrt{2^n-m}$ and $f(1)=\sin((2p+1)\theta)/\sqrt{m}$, where $\theta=\sin^{-1}\sqrt{m/2^n}$ and $m$ is the degeneracy of $w=1$. Equation~\eqref{eqn: psivar} generalizes this result to arbitrary diagonal observables $\hW$ with an arbitrary number of eigenvalues.

\begin{figure}[t]
\includegraphics[width=0.95\columnwidth]{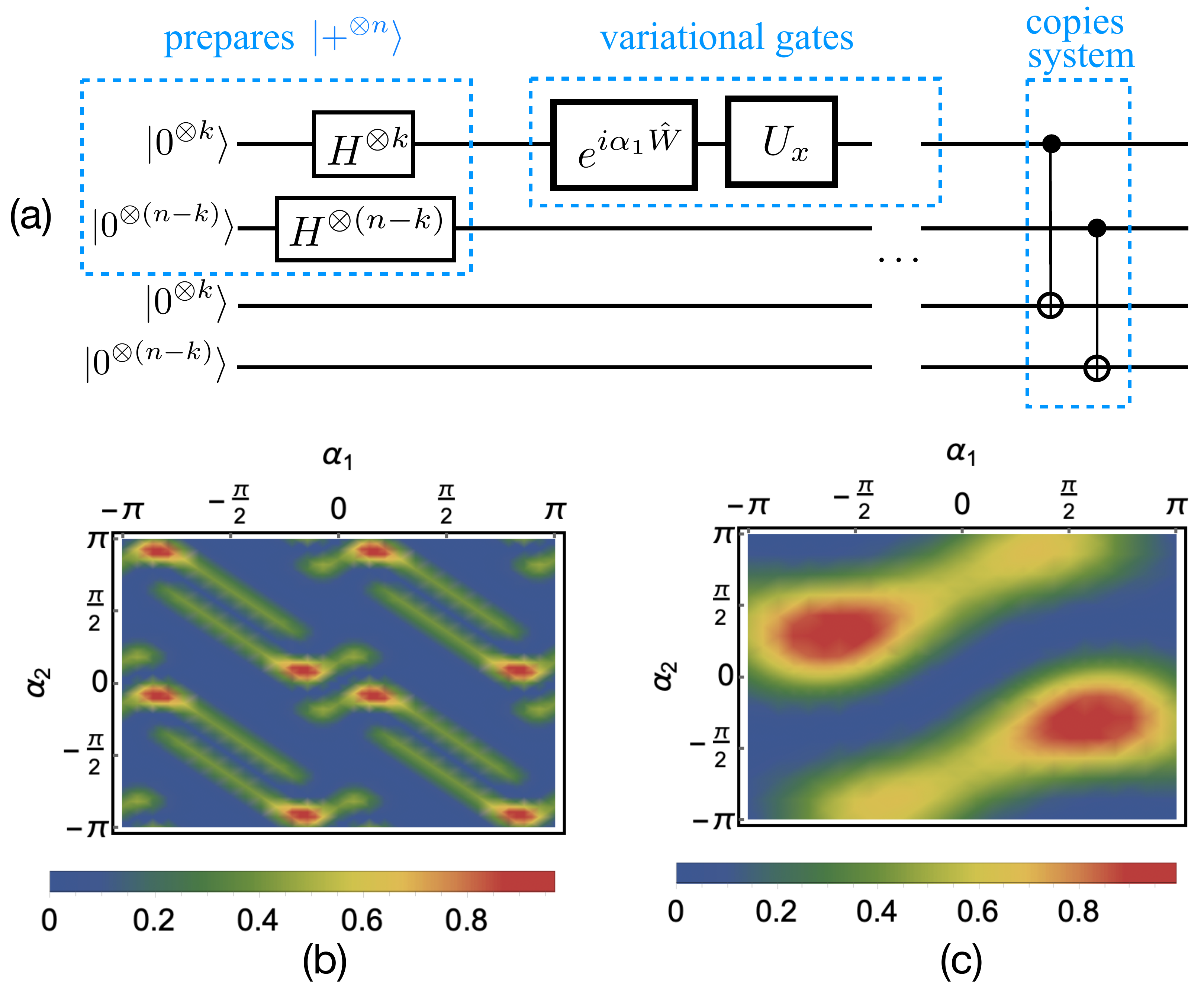}
\caption{(Color online) (a) Quantum circuit that prepares the variational ansatz [Eq.~\eqref{eqn: psivar}] for $\ket{W_1}$ [Eq.~\eqref{eqn: W1}]. H refers to the Hadamard gate, and the last gate on the right is bitwise CNOT. (b) Magnitude of the fidelities of the variational ansatz [Eq.~\eqref{eqn: psivar}] at $p=2$, with $\ket{W_1}$ [Eq.~\eqref{eqn: W1}], when (a) $\hW=\sum_{i=1}^5 \hs^z_i$, and (b) $\hW$ has a uniform eigenvalue distribution in $[-1,1]$. The maximum fidelity for these two cases is respectively $0.994$ and $0.999$.}
\label{fig4: purified state prep}
\end{figure}

We use $\ket{\psi_{\rm var}}$ as a variational ansatz for $\ket{W_1}$, with variational parameters $\alpha_{1\cdots p}$ that are chosen to maximize the fidelity of $\ket{\psi_{\rm var}}$ with $\ket{W_1}$, for a given $p$. The optimal fidelity increases with $p$, and reaches $1$ if $f(w)=w/\sqrt{\tr(\hW\hW\+)}$. Figure~\ref{fig4: purified state prep}(a) shows the quantum circuit to prepare $\ket{\psi_{\rm var}}$.

Crucial to our protocol is that the formulae for the fidelities $F_p(\alpha_{1\cdots p})$ for given $\alpha_{1\cdots p}$ can be found straightforwardly and analytically, in terms of only the eigenvalues of $\hW$. We analytically derive the fidelities in the Supplementary Material. Remarkably, we find that for a wide range of $\hW$ with different eigen spectra, the maximum value of $F_2$ is greater than $0.99$.

Figures~\ref{fig4: purified state prep}(b-c) plot the fidelity $F_2(\alpha_1,\alpha_2)$ for two different $\hW$. Figures~\ref{fig4: purified state prep}(b) considers $\hW = \sum_{i=1}^5\hs^z_i$, and Fig.~\ref{fig4: purified state prep}(c) considers $\hW$ with a uniform eigenvalue distribution. We find that the maximum of $F_2$ is greater than $0.99$ in both cases. When $\hW$ is a Pauli operator, the maximum fidelity at $p=1$ is $F_1(\alpha_1=\pi/2)=1$. In this case, the circuit in Fig.~\ref{fig4: purified state prep}(a) reduces to Fig.~\ref{fig1: definition}(b).

Implementation of $U_x$ in Fig.~\ref{fig4: purified state prep}(a) on digital quantum platforms using only one- and two-qubit gates is cumbersome, but well known from the literature on Grover's algorithm~\cite{barenco1995elementary, saeedi2013linear}. Alternatively, $U_x$ can be directly implemented using the Rydberg blockade, without being deconstructed into two-qubit gates~\cite{muller2009mesoscopic, su2018one, brion2007conditional, wu2010implementation, isenhower2011multibit, saffman2009efficient, wu2017rydberg, young2020asymmetric}.

In conclusion, we showed how to measure the infinite-temperature OTOC by making simple measurements on two systems that are initially entangled and then evolved with the Hamiltonian, for Hamiltonians with either a chiral symmetry or a particle-hole symmetry. We showed that the initial state is a product of Bell pairs for unitary $\hW$, produced a variational ansatz for non-unitary easily diagonalizable $\hW$, and analytically derived the fidelity of this ansatz with the desired initial state. Although our method works for a restricted class of Hamiltonians, and requires twice the number of qubits, it has some benefits -- it can be implemented without reversing the Hamiltonian's sign, requires fewer measurements than some earlier methods, and is robust to errors. Our ideas can be experimentally implemented on currently available quantum platforms.

Our method may also complement schemes that implement the SWAP operator using randomized measurements or initial states~\cite{brydges2019probing, elben2019statistical, elben2020many, elben2020cross, cian2020many}, to extract other physical quantities using fewer measurements or with smaller statistical errors. The ability to effectively evolve one half of Bell pairs with $-\hH$ by evolving the other half with $+\hH$, for a restricted family of Hamiltonians, may have applications in scenarios that involve quantum echoes.

\section*{Acknowledgment}
We thank R. van Bijnen and M. K. Joshi for valuable discussions, A. Kruckenhauser for valuable discussions and comments on the manuscript, and A. Elben, L. K. Joshi, D. Vasilyev and P. Zoller for valuable discussions and critical remarks on the manuscript. B.S. acknowledges funding from the European Union's Horizon 2020 research and innovation programme under Grant Agreement No. 817482 (Pasquans) and No. 731473 (QuantERA via QT-FLAG). Furthermore, this work was supported by the Simons Collaboration on Ultra-Quantum Matter, which is a grant from the Simons Foundation (651440, P.Z.), and LASCEM by AFOSR No. 64896-PH-QC.

\bibliography{refs}

\renewcommand{\theequation}{S\arabic{equation}}
\renewcommand{\thefigure}{S\arabic{figure}}
\setcounter{figure}{0} 
\setcounter{equation}{0} 

\section{Appendix A: Effects of decoherence and imperfections on OTOC measurements}
Here we give further details about the simulations of errors expected to occur in an experimental implementation of our proposal. Our goal is to explore the robustness of the rescaled OTOC, $\overline{O}(t) = \langle W_{12}(t)| \hV\otimes\hV^T |W_{12}(t)\rangle / \langle {\rm Bell}(t)| \hV\otimes\hV^T |{\rm Bell}(t)\rangle$, to these errors. For simplicity and concreteness, we consider $\hV = \hs^x_j$ and $\hW = \hs^z_i$ throughout this section. We set $\hbar=1$.

\subsection{A.I: Readout errors}
Our protocol measures $\hV\otimes \hV^T$ after evolving the system. Let the ideal probabilities of measuring $\hV \otimes \hV^T$ as $(\pm 1,\pm 1)$ be $P_{\pm 1,\pm 1}$ in the limit of $N_m=\infty$ measurements, and the error probability for each measurement be $x$. The ideal expectation value of $\hV\otimes \hV^T$ is
\begin{equation}
\langle \hV\otimes \hV^T \rangle = \sum_{\sigma_1,\sigma_2=\pm1} \sigma_1\sigma_2 P_{\sigma1,\sigma2}.
\end{equation}
However, due to readout errors, the actual probabilities for measuring $(\pm1,\pm1)$ are
\begin{align}
P^{\rm est}_{\sigma_1,\sigma_2} = &(1-x)^2P_{\sigma_1,\sigma_2} + x(1-x)(P_{\sigma_1,-\sigma_2} + P_{-\sigma_1,\sigma_2})\nonumber\\ & + x^2 P_{-\sigma_1,-\sigma_2}
\end{align}
giving the estimated expectation value for $\langle \hV\otimes \hV^T \rangle$ as
\begin{align}
\langle \hV\otimes \hV^T \rangle^{\rm est}  =& \sum_{\sigma_1,\sigma_2=\pm1} \sigma_1\sigma_2 P^{\rm est}_{\sigma1,\sigma2} \nonumber\\
=& \sum_{\sigma_1,\sigma_2=\pm1} \sigma_1\sigma_2 \big( (1-x)^2 P^{\rm est}_{\sigma_1,\sigma_2}\nonumber\\ & + x(1-x) (P^{\rm est}_{\sigma_1,-\sigma_2} + P^{\rm est}_{-\sigma_1,\sigma_2}) + x^2 P^{\rm est}_{-\sigma_1,-\sigma_2} \big)\nonumber\\
=& (1-2x)^2 \sum_{\sigma_1,\sigma_2=\pm1} \sigma_1\sigma_2 P_{\sigma1,\sigma2}\nonumber\\
=& (1-2x)^2 \langle \hV\otimes \hV^T \rangle.
\end{align}
Thus, readout errors rescale the ideal expectation value by $(1-2x)^2$. This rescaling occurs both in $O^{\rm est}(t)$ and in $O'(t)$, and the ratio of these two quantities $\overline{O}(t) = O^{\rm est}(t)/O'(t)$ is left unchanged.

\subsection{A.II: Errors in the initial state}
The ideal initial state in our protocol, for the example we considered, is a product of Bell pairs, $\hat{\rho}_{\rm init} = \otimes_j \hat{\rho}_j$ with  $\hat{\rho}_j = \ket{\Phi^\pm_j}\bra{\Phi^\pm_j}$ giving the appropriate Bell state for the $j^{\rm th}$ qubit. We model imperfect initial state preparation by writing the initial density matrix as $\hat{\rho}_{\rm init} = \otimes_j ((1-\delta) \hat{\rho}_j + \hat{1}\times\delta/4)$. Here, the fidelity to prepare each Bell pair is $1-\delta$. Current experiments have $1-\delta
\sim0.98$.

A measurement of $\hV\otimes\hV^T$ at time $t$ is affected only by the imperfect Bell pairs that lie within the support of $\hV(t)$. At $t=0$, $\hV$ has support on only one qubit, and the imperfect state preparation rescales the measurement by $(1-\delta)$ to produce $O_{ij}^{\rm est}(t) = (1-\delta)O_{ij}(t)$. This error is exactly cancelled in the ratio $\overline{O}_{ij}(t) = O^{\rm est}_{ij}(t)/O'(t)$. As the operator spreads with time, the error in $O^{\rm est}_{ij}(t)$ and $O'(t)$ grow, but these are also cancelled in $\overline{O}_{ij}(t)$ as long as $\hV(t)$ has not yet spread to the location of $\hW$. The cancellation is not exact after $\hV(t)$ reaches $\hW$.

Figure~\ref{fig1: supp}(a-b) plot the error in $O_{1n}^{\rm est}(t), O'_n(t)$ and $\overline{O}_{1n}(t)$ due to imperfect state preparation. Figure~\ref{fig1: supp}(a) shows that the errors increase linearly with $\delta$ for a fixed time, and that $|O_{1n}(t)-\overline{O}_{1n}(t)| < |O_{1n}(t)-O_{1n}^{\rm est}(t)|$. Figure~\ref{fig1: supp}(b) shows that the errors in $O_{1n}^{\rm est}(t)$ and $O'_n(t)$ begin increasing linearly from $t=0$, but the error in $\overline{O}_{1n}(t)$ stays close to $0$ until $\hs^x_n(t)$ spreads to qubit $1$ at $Jt\sim1$. For $Jt>1$, the error in $\overline{O}_{1n}(t)$ also increases linearly with time.

We note that in systems with chaos (which is not exhibited in our case), the support of $\hV(t)$ grows exponentially with time. Then the error in the measured OTOC is also expected to grow exponentially, consistent with the usual arguments of sensitivity to initial conditions in chaotic systems.

\subsection{A.III: Symmetry-breaking terms in the Hamiltonian}
\begin{figure}[t]
\includegraphics[width=1.0\columnwidth]{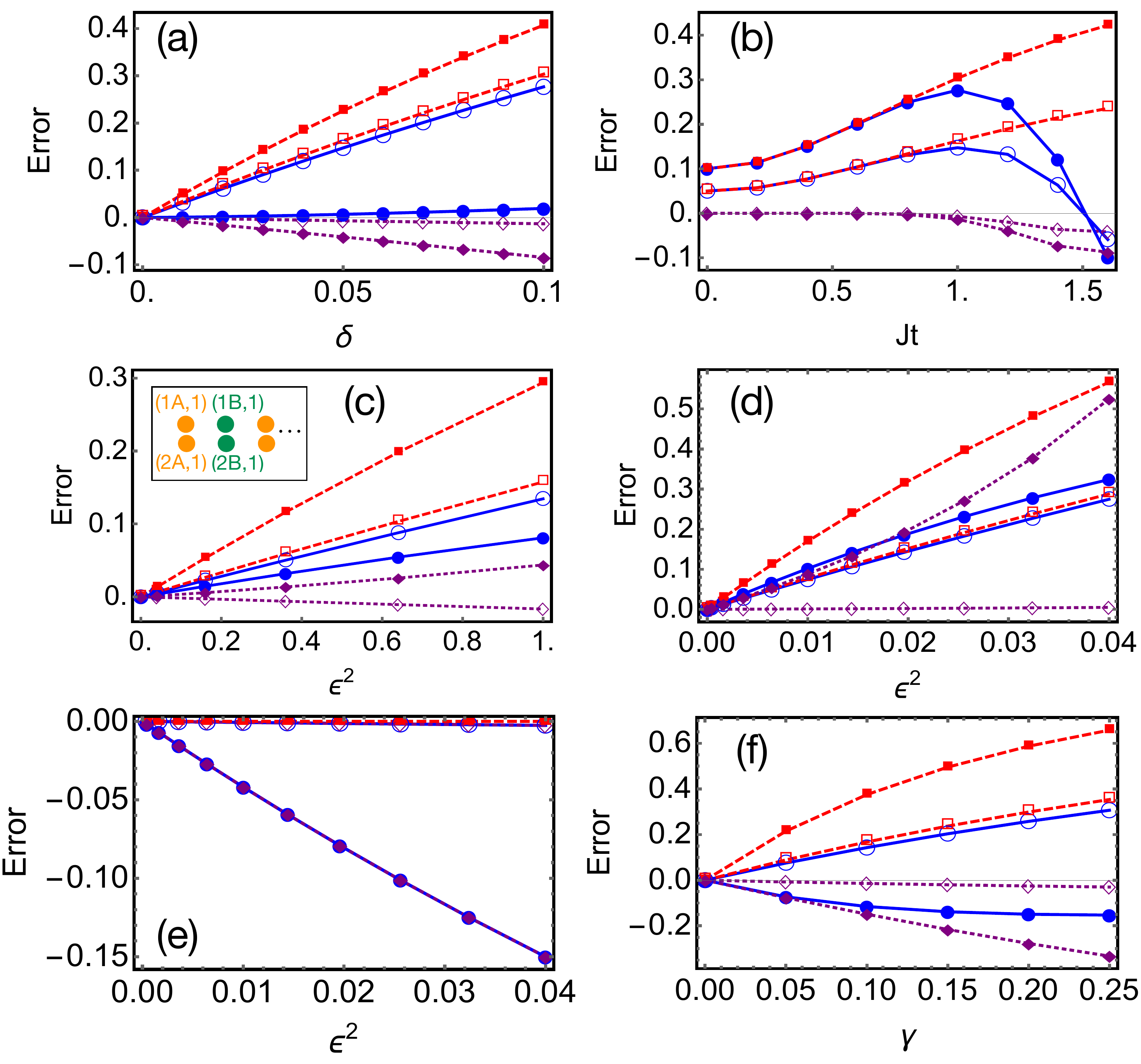}
\caption{ (Color online) Errors in OTOC measurements for varying strengths of imperfections and times. (a-b) The initial state is taken as $\hat{\rho}_{\rm init} = \otimes_j ((1-\delta) \hat{\rho}_j + \hat{1}\times \delta/4)$, with fidelity $1-\delta$ for preparing each Bell pair. (c) The Hamiltonian for each system is $\hH_\epsilon = \hH_{AB} + \epsilon(\hH_{AA} + \hH_{BB})$ [see Eq.~\eqref{eqn: HAB}], where $\hH_{AA}$ and $\hH_{BB}$ break chiral symmetry. (d) The two systems evolve with unequal Hamiltonians $\hH_{1(2)}$ [see Eq.~\eqref{eqn: Hunequal}].  (e) The two systems have the total Hamiltonian $\hH_\epsilon = \hH_{1A,1B} + \hH_{2A,2B} + \epsilon(\hH_{1A,2A} + \hH_{1B,2B})$ [see Eq.~\eqref{eqn: H12}], where $\hH_{1A,2A}$ and $\hH_{1B,2B}$ couple systems 1 and 2. (f) We include spontaneous emission with decay rate $\gamma$ for each qubit. Blue circles connected by solid lines plot $O_{1n}(t)-O_{1n}^{\rm est}(t)$, red squares connected by dashed lines plot $1-O'_n(t)$, and purple diamonds connected by dotted lines plot $O_{1n}(t)-\overline{O}_{1n}(t)$. In (b), the open and closed symbols correspond to $\delta=0.05$ and $\delta=0.1$, respectively. In (a,c-e), open and closed symbols correspond to $Jt=1$ and $Jt=1.5$, respectively. In (f), open and closed symbols correspond to $Jt=0.8$ and $Jt=1.3$, respectively. The number of qubits is $n=6$ in (a)-(e), and $n=5$ in (f). Inset in (c) shows a schematic of the two systems in consideration.
}
\label{fig1: supp}
\end{figure}

To illustrate robustness to breaking of chiral symmetry or particle-hole symmetry, we consider the system to have the Hamiltonian
\begin{align}\label{eqn: HAB}
& \hH_\epsilon = \hH_{AB} + \epsilon(\hH_{AA} + \hH_{BB}), \nonumber\\
& \hH_{AB} = \sum_{ij} \frac{J}{r_{ij}^3} \left(\hs^x_{A,i}\hs^x_{B,j} + \hs^y_{A,i}\hs^y_{B,j} \right), \nonumber\\
& \hH_{AA} = \sum_{ij} \frac{J}{r_{ij}^3} \left(\hs^x_{A,i}\hs^x_{A,j} + \hs^y_{A,i}\hs^y_{A,j} \right), \nonumber\\
& \hH_{BB} = \sum_{ij} \frac{J}{r_{ij}^3} \left(\hs^x_{B,i}\hs^x_{B,j} + \hs^y_{B,i}\hs^y_{B,j} \right).
\end{align}
We assume an equally spaced linear chain of qubits which is bipartitioned into A and B sublattices, as shown in the inset of Fig.~\ref{fig1: supp}(c). We take the nearest-neighbor distance to be $1$. The experiment~\cite{de2019observation} which realized a variant of $\hH_{AB}$ had a $2\times(n/2)$ ladder of atoms and not a linear chain, but we consider a linear chain here for simplicity. We group the intra-sublattice terms as $\hH' = \hH_{AA} + \hH_{BB}$. These terms may arise if the two legs of the ladder in the experiment are not aligned properly. 

$\hH_{AB}$ has chiral symmetry, and $\hH'$ breaks this symmetry. In the basis chosen in the main text, $(\hH')^T=+\hH'$. Any other perturbations to the Hamiltonian, if present, can also be separated into terms that have a symmetric matrix (i.e. real matrix elements), and terms that have an anti-symmetric matrix (i.e imaginary matrix elements). All terms with real matrix elements break chiral symmetry.

We denote
\begin{align}
O_\epsilon(t) =& \bra{W_{12}} \left(e^{i\hH_\epsilon t} \hV e^{-i\hH_\epsilon t}\right) \otimes \left(e^{i\hH_\epsilon t} \hV^T e^{-i\hH_\epsilon t}\right) \ket{W_{12}}.
\end{align}
A similar derivation to Eq.~(5) in the main text yields
\begin{align}\label{eqn: temp1}
O_\epsilon(t) =& \frac{ 1 }{ 2^n } \tr(\hW e^{i(\hH+\epsilon\hH')t}\hV e^{-i(\hH+\epsilon\hH')t}\nonumber\\ &\times \hW e^{i(\hH-\epsilon\hH')t} \hV e^{-i(\hH-\epsilon\hH')t} ).
\end{align}
Due to the cyclic property of the trace, Eq.~\eqref{eqn: temp1} can be reorganized as
\begin{align}\label{eqn: temp2}
O_\epsilon(t) =& \frac{ 1 }{ 2^n } \tr( \hW e^{i(\hH-\epsilon\hH')t}\hV e^{-i(\hH-\epsilon\hH')t}\nonumber\\ &\times \hW e^{i(\hH+\epsilon\hH')t} \hV e^{-i(\hH+\epsilon\hH')t} ) \nonumber\\
=& O_{-\epsilon}(t).
\end{align}
Thus $O_\epsilon(t)$ is an even function of $\epsilon$, and therefore the error in $O_\epsilon(t)$ scales as $\epsilon^2$ at leading order. A similar argument can be made for the scaling of $O'(t)$.

Figure~\ref{fig1: supp}(c) confirms the argument above, for a particular pair $(i,j)=(1,n)$.

\subsection{A.IV: Evolution with unequal Hamiltonians}
To illustrate robustness to evolution with unequal Hamiltonians, we consider the two systems to have the Hamiltonians
\begin{equation}\label{eqn: Hunequal}
 \hH_{1(2)} = (1+\epsilon_{1(2)}) \hH_{AB},
\end{equation}
with $\epsilon_{1(2)} = +\epsilon(-\epsilon)$. This asymmetry between the two systems may arise if the particles in the two systems interact with different strengths.

We denote
\begin{align}
O_\epsilon(t) = &\bra{W_{12}} \left(e^{i(1+\epsilon)\hH t} \hV e^{-i(1+\epsilon)\hH t}\right)\nonumber\\ &\otimes \left(e^{i(1-\epsilon)\hH t} \hV^T e^{-i(1-\epsilon)\hH t}\right) \ket{W_{12}}
\end{align}
Using the same arguments of trace-cyclicality as before [see Eqs.~\eqref{eqn: temp1} and~\eqref{eqn: temp2}], we find $O_\epsilon(t) = O_{-\epsilon}(t)$. Therefore, therefore the error in $O_\epsilon(t)$ scales as $\epsilon^2$ at leading order. A similar argument can be made for the scaling of $O'(t)$ with $\epsilon$.

Figure~\ref{fig1: supp}(d) confirms the argument above, for $(i,j)=(1,n)$.

\subsection{A.V: Coupling between the two systems}
To illustrate robustness to coupling between systems 1 and 2, we consider the two systems to have the total Hamiltonian
\begin{align}\label{eqn: H12}
& \hH_\epsilon = \hH_{1A,1B} + \hH_{2A,2B} + \epsilon(\hH_{1A,2A} + \hH_{1B,2B}), \nonumber\\
& \hH_{kA,kB} = \sum_{ij} \frac{J}{r_{ij}^3} \left(\hs^x_{kA,i}\hs^x_{kB,j} + \hs^y_{kA,i}\hs^y_{kB,j} \right), \nonumber\\
& \hH_{1A,2A} = \sum_i J \left(\hs^x_{1A,i}\hs^x_{2A,i} + \hs^y_{1A,i}\hs^y_{2A,i} \right), \nonumber\\
& \hH_{1B,2B} = \sum_i J \left(\hs^x_{1B,i}\hs^x_{2B,i} + \hs^y_{1B,i}\hs^y_{2B,i} \right).
\end{align}
Here, 1A(2A) and 1B(2B) denote the A and B sublattices in system 1(2).

We denote
\begin{align}
O_\epsilon(t) =& \bra{W_{12}} \left(e^{i\hH_\epsilon t}\hV e^{-i\hH_\epsilon t}\right)\otimes  \left(e^{i\hH_\epsilon t}\hV^Te^{-i\hH_\epsilon t}\right) \ket{W_{12}} \nonumber\\
=&  \bra{\rm Bell} \left(\hW\+e^{i\hH_\epsilon t}\hV e^{-i\hH_\epsilon t}\hW\right)\nonumber\\ &\otimes  \left(e^{i\hH_\epsilon t}\hV^Te^{-i\hH_\epsilon t}\right) \ket{{\rm Bell}}
\end{align}
In the basis chosen in the main text, $\hW$ and $\hV$ have real matrices, the unperturbed Hamiltonians $\hH_{1A,1B}$ and $\hH_{2A,2B}$ have imaginary matrices, while the coupling terms $\hH_{1A,2A}$ and $\hH_{1B,2B}$ have real matrices. By definition, $\ket{{\rm Bell}}$ is a real vector. Then, using $\hH_\epsilon^* = -\hH_{-\epsilon}$, we find
\begin{align}
O_\epsilon^*(t) =& \bra{\rm Bell} \left(\hW\+e^{-i\hH_\epsilon^* t}\hV e^{i\hH_\epsilon^* t}\hW\right) \nonumber\\ &\otimes  \left(e^{-i\hH_\epsilon^* t}\hV^Te^{i\hH_\epsilon^* t}\right) \ket{{\rm Bell}} \nonumber\\
=& O_{-\epsilon}(t)
\end{align}
Since $O_\epsilon(t)$ is real, we find that $O_\epsilon(t)$ is an even function of $\epsilon$, implying that the error in $O_\epsilon(t)$ scales as $\epsilon^2$ at leading order.

Figure~\ref{fig1: supp}(e) confirms the argument above, for $(i,j)=(1,n)$.

For the coupling considered above, $\ket{{\rm Bell}}$ is an eigenstate of $e^{-i\hH_\epsilon t} \otimes e^{-i\hH_\epsilon t}$ and an eigenstate of $\hs^x_j\otimes(\hs^x_j)^T$. Therefore, $O'_j(t) = 1$ even in the presence of coupling between systems 1 and 2, as illustrated by the red dashed line in Fig.~\ref{fig1: supp}(d). Then, unlike all the other errors above that can be detected by measuring $O'_j(t)$, the presence of this coupling between systems 1 and 2 will not be revealed by measuring $O'_j(t)$.

\subsection{Appendix A.VI: Spontaneous emission}
We include spontaneous emission via the Lindblad master equation
\begin{align}
&\frac{ \partial\hat\rho }{ \partial t} = -i[\hH, \hat\rho(t)] + \gamma \sum_j \mathcal{L}[\hs^-_j](\hat\rho(t)), \nonumber\\
&\mathcal{L}[\hs^-_j](\hat\rho) = \frac{1}{2} (2\hs^-_j\hat\rho\hs^+_j - \hat\rho\hs^+_j\hs^-_j - \hs^+_j\hs^-_j\hat\rho),
\end{align}
with the initial condition $\hat\rho(0) = \ket{W_{12}}\bra{W_{12}}$, and $\gamma$ the spontaneous emission rate (identical for each qubit).

Figure~\ref{fig1: supp}(f) plots the errors in $O_{ij}^{\rm est}(t), O'_j(t)$ and $\overline{O}_{ij}(t)$ as a function of $\gamma$.

\subsection{Appendix A.VII: Depolarizing noise}
Depolarizing noise is included via the Lindblad master equation
\begin{equation}\label{eqn: depolarizing noise Lindblad}
\frac{ \partial\hat\rho }{ \partial t} = -i[\hH, \hat\rho(t)] + \gamma (\frac{\hat{1}}{4^n} - \hat\rho )
\end{equation}
with the initial condition is $\hat\rho(0) = \ket{W_{12}}\bra{W_{12}}$, $\gamma$ the depolarizarion rate, and $\hH$ the total Hamiltonian for the two systems combined. The solution to Eq.~\eqref{eqn: depolarizing noise Lindblad} is
\begin{equation} \label{eqn: depolarizing noise}
\hat\rho(t) = e^{-\gamma t} e^{-i\hH t}\hat\rho(0) e^{i\hH t} + (1-e^{-\gamma t})\hat{1}/4^n
\end{equation}

The expectation value of $\hV\otimes \hV^T$ in $\rho(t)$ is
\begin{equation}\label{eqn: depolarizing noise rescaling}
\langle \hV\otimes\hV^T\rangle_\infty = e^{-\gamma t} \tr(\hV(t)\hV^T(t)\rho(0))
\end{equation}
since the second term in Eq.~\eqref{eqn: depolarizing noise} does not contribute. The right hand side of Eq.~\eqref{eqn: depolarizing noise rescaling} is $e^{-\gamma t}$ times the ideal measurement without errors. That is, depolarizing noise rescales both $O_{ij}^{\rm est}(t)$ and $O'_j(t)$ by $e^{-\gamma t}$. The factor $e^{-\gamma t}$ is cancelled in the ratio $\overline{O}_{ij}(t) = O_{ij}^{\rm est}(t)/O'_j(t)$, and therefore the error in the rescaled OTOC is zero.

\section{Appendix C: Proof of Eq.~(8) in the main text}

\begin{figure}[t]
\includegraphics[width=1.0\columnwidth]{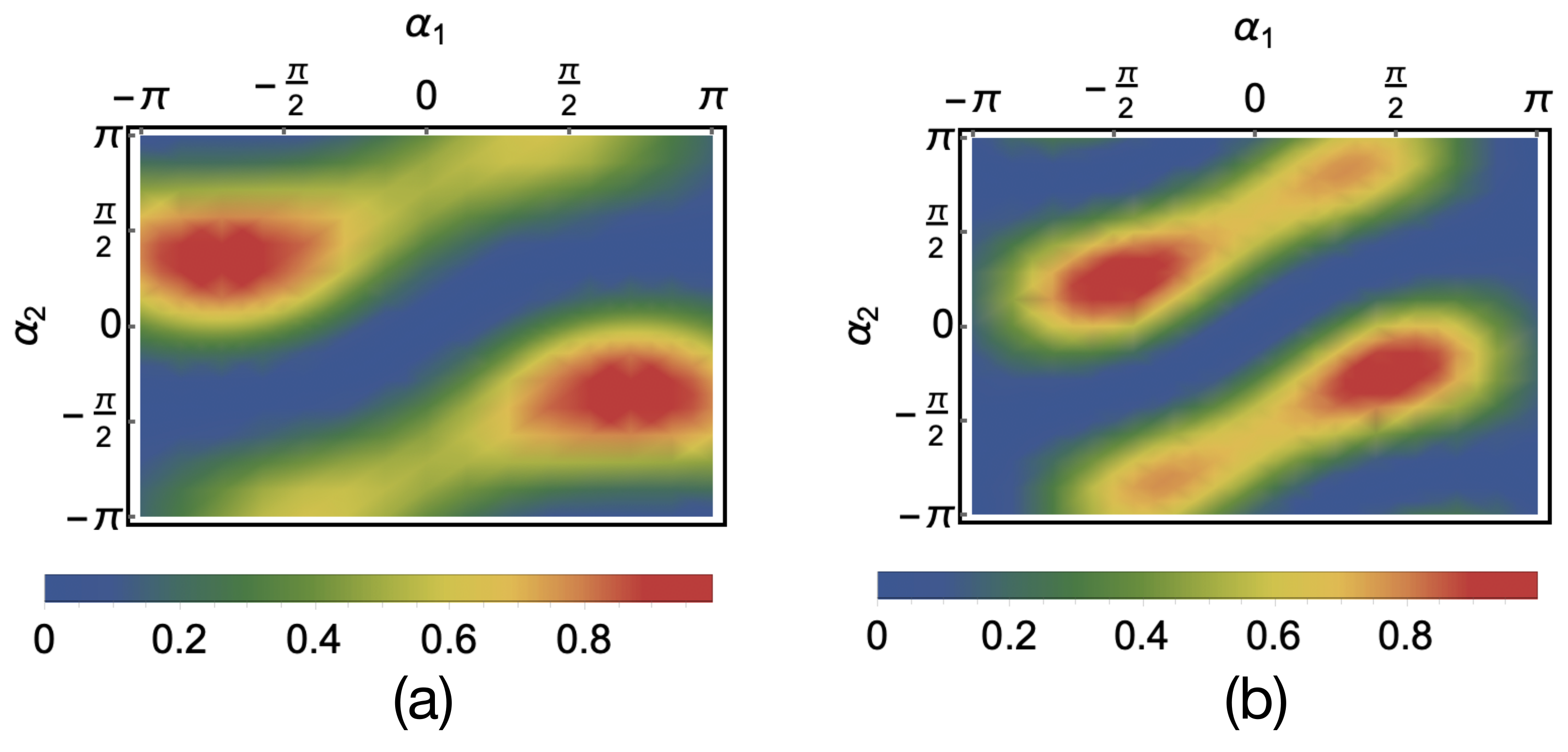}
\caption{(Color online) Magnitude of the fidelities of the variational ansatz $\ket{\psi_{\rm var}}$ at $p=2$, with the target state $\ket{W_1}$, when the eigenvalue distribution for $\hW$ is given by (a) a Wigner semicircle distribution, and (b) the arcsine distribution. The maximum fidelity at $p=2$ for these two cases is respectively $0.999$ and $0.9997$.}
\label{fig1: fidelities}
\end{figure}

\begin{table}[t]
\begin{tabular}{|c|ccc|c|}
\hline
Eigenvalue distribution of $\hW$ && $p$ && Maximum fidelity\\
\hline
Bernoulli ($q=0.5$) && 1 && 1\\
Arcsine && 2 && 0.9997\\
Wigner semicircle && 2 && 0.999\\
Uniform && 2 && 0.999\\
Gaussian && 2 && 0.991\\
\hline
\end{tabular}
\caption{Maximum fidelities for $\ket{\psi_{\rm var}}$ with $\ket{W_1}$, for different eigenvalue distributions for $\hW$ and low variational depth.}
\label{table}
\end{table}

Suppose $\hW$ has distinct eigenvalues $w_0, w_1, \cdots$, with degeneracies $\mathcal{N}(w_0), \mathcal{N}(w_1), \cdots$. For each \textit{distinct} eigenvalue $w_i$, define the manifold of states with that eigenvalue as $S(w_i)$, and 
\begin{equation}
\ket{\Phi(w_i)} = \frac{1}{\sqrt{\mathcal{N}(w_i)}} \sum_{\phi\in S(w_i)} \ket{\phi}.
\end{equation}
This defines a one-to-one map between the set $\{\ket{\Phi(w_i)}\}$ and the set of distinct eigenvalues $\{w_i\}$. The states $\{\ket{\Phi(w_i)}\}$ are orthogonal to each other, and form a complete basis for a subspace of the full Hilbert space.

The main argument of our proof below is that the variational ansatz in Eq.~(8) lies in the Hilbert space spanned by $\{\ket{\Phi(w_i)}\}$, and therefore can be written as a superposition of only $\{\ket{\Phi(w_i)}\}$. The coefficients in this superposition are $f(w_i)\sqrt{\mathcal{N}(w_i)}$.

Since $\hW$ is diagonal, its eigenstates $\ket{\phi}$ are bit strings. We label the qubits where $\hW$ has support to be $[1,k]$. Then, the eigenstates $(x_1\cdots x_k \cdots 0_j \cdots)$ and $(x_1\cdots x_k \cdots 1_j \cdots)$ are degenerate, and will appear in $\ket{\Phi(w_i)}$ in equal superposition. Extending this argument to all qubits outside $[1,k]$, we find that $\ket{\Phi(w_i)}$ can be factorized as
\begin{equation}
\ket{\Phi(w_i)} = \left(\frac{1}{\sqrt{\mathcal{N}_k(w_i)}} \sum_{\phi\in S_k(w_i)} \ket{\phi}\right) \ket{+^{\otimes(n-k)}},
\end{equation}
where $S_k(w_i)$ is the set of bit strings of length $k$ with eigenvalue $w_i$, and $\mathcal{N}_k(w_i) = \mathcal{N}(w_i)/2^{n-k}$. We denote $\ket{\Phi_k(w_i)} =  \sum_{\phi\in S_k(w_i)} \ket{\phi}/\sqrt{\mathcal{N}_k(w_i)}$. Then, $\ket{\Phi(w_i)} = \ket{\Phi_k(w_i)}\ket{+^{\otimes(n-k)}}$.

The left hand side of Eq.~(8) factorizes as
$$\left( \prod_{j=1}^p U_x e^{i\alpha_j\hW} \ket{+^{\otimes k}} \right) \ket{+^{\otimes(n-k)}},$$
since $U_x$ and $\hW$ act only on the first $k$ qubits. The right hand side of Eq.~(8) factorizes as
$$\left( \sum_{w_i} f(w_i) \sqrt{\mathcal{N}(w_i)} \ket{\Phi_k(w_i)}\right) \ket{+^{\otimes(n-k)}},$$
where the sum runs over the \textit{distinct} eigenvalues of $\hW$.

Thus, to prove Eq.~(8) in the main text, it is sufficient to prove that 
$ \prod_{j=1}^p U_x e^{i\alpha_j\hW} \ket{+^{\otimes k}}$ 
lies in the Hilbert space spanned by $\{\ket{\Phi_k(w_i)}\}$. We prove this by showing that $\ket{+^{\otimes k}}$ lies in this Hilbert space, and both $U_x$ and $\hW$ are closed under this Hilbert space.

$\ket{\Phi_k(w_i)}$ is an eigenstate of $\hW$ with eigenvalue $w_i$. Therefore, $\hW$ is closed under this Hilbert space.

We can write
\begin{equation} \label{eqn: psi0}
\ket{+^{\otimes k}} = \frac{1}{\sqrt{2^k}}\sum_{\phi=0}^{2^k-1} \ket{\phi} = \sum_{w_i} \sqrt{\frac{\mathcal{N}(w_i)}{2^n}} \ket{\Phi_k(w_i)},
\end{equation}
where the first sum runs over all the eigenstates of $\hW$, and the second sum runs over only the \textit{distinct} eigenvalues $w_i$. Therefore, $\ket{+^{\otimes k}}$ lies in the Hilbert space spanned by $\{\ket{\Phi_k(w_i)}\}$. Then, $U_x=1-2\ket{+^{\otimes k}}\bra{+^{\otimes k}}$ is also closed under this Hilbert space. This completes the proof of Eq.~(8).

The target state,
\begin{equation} \label{eqn: target}
\ket{W_1} = \left( \sum_{w_i} w_i\sqrt{\frac{\mathcal{N}(w_i)}{\tr(\hW\hW\+)}} \ket{\Phi_k(w_i)}\right)  \ket{+^{\otimes(n-k)}},
\end{equation}
is equal to Eq.~(8) when $f(w_i) =  w_i/\sqrt{\tr(\hW\hW\+)}$.

\section{Appendix D: Calculating the fidelities}
The fidelity of $\ket{+^{\otimes n}}$ with $\ket{W_1}$ is (using Eqs.~\eqref{eqn: target} and~\eqref{eqn: psi0})
\begin{align}
F_0 =& \overlap{W_1}+^{\otimes n}\nonumber\\
 = &\sum_{w_i} \sqrt{\frac{\mathcal{N}(w_i)}{2^n}} \times w_i \sqrt{\frac{\mathcal{N}(w_i)}{\tr(\hW\hW\+)}} \overlap{\Phi_k(w_i)}{\Phi_k(w_i)} \nonumber\\
=& \sum_{w_i} \frac{w_i\mathcal{N}(w_i)}{\sqrt{2^n\tr(\hW\hW\+)}} = \frac{\tr(\hW)}{\sqrt{2^n\tr(\hW\hW\+)}}.
\end{align}
To derive the last equality in this equation, we used the relation $\sum_{w_i} g(w_i) \mathcal{N}(w_i) = \tr(g(\hW))$ for any function $g$.

The fidelity of $\ket{\psi_{\rm var}(\alpha)}$ with $\ket{W_1}$ is
\begin{align}
F_1(\alpha) =& \overlap{W_1}{\psi_{\rm var}(\alpha)}\nonumber\\
=& \bra{W_1} (1-2\ket{+^{\otimes n}}\bra{+^{\otimes n}}) e^{i\alpha\hW} \ket{+^{\otimes n}} \nonumber\\
=&  \bra{W_1}e^{i\alpha\hW}\ket{+^{\otimes n}} - 2 \overlap{W_1}{+^{\otimes n}} \bra{+^{\otimes n}}e^{i\alpha\hW}\ket{+^{\otimes n}} \nonumber\\
=& \left( \sum_{w_i} \frac{w_i e^{i\alpha w_i} \mathcal{N}(w_i)}{\sqrt{2^n\tr(\hW\hW\+)}}  \right) - 2 \bigg(\frac{\tr(\hW)}{\sqrt{2^n\tr(\hW\hW\+)}} \nonumber\\ &\times \sum_{w_i} \frac{e^{i\alpha w_i} \mathcal{N}(w_i)}{2^n}\bigg)\nonumber\\
=& \frac{ \tr(\hW e^{i\alpha\hW}) - 2\tr(\hW)\tr(e^{i\alpha\hW})/2^n } {\sqrt{2^n\tr(\hW\hW\+)}}.
\end{align}

For $p>1$, the fidelity can be calculated recursively as
\begin{align}
F_p(\alpha_1\cdots\alpha_p) = & \bra{W_1} \cdots e^{i\alpha_2\hW} U_x e^{i\alpha_1\hW} \ket{+^{\otimes n}} \nonumber\\
=& \bra{W_1} \cdots e^{i\alpha_2\hW} e^{i\alpha_1\hW} \ket{+^{\otimes n}} \nonumber\\ &- 2 \bra{W_1} \cdots e^{i\alpha_2\hW} \ket{+^{\otimes n}} \bra{+^{\otimes n}} e^{i\alpha_1\hW} \ket{+^{\otimes n}} \nonumber\\
=& F_{p-1}(\alpha_1+\alpha_2,\alpha_3\cdots\alpha_p) \nonumber\\
 & - 2F_{p-1}(\alpha_2\cdots\alpha_p) \frac{\tr(e^{i\alpha_1\hW})}{2^n}.
\end{align}

Figure~\ref{fig1: fidelities} plots $F_2$ for $\hW$ with eigenvalue distributions given by the Wigner semicircle distribution and the arcsine distribution. These distributions are chosen for illustrative purposes. We find that the maximum of $F_2$ is greater than $0.999$ in both cases. Table~\ref{table} lists the maximum fidelities for five eigenvalue distributions of $\hW$.

\end{document}